\documentclass[aip,jcp,graphicx,reprint,flotfix]{revtex4-1}

\usepackage{inputenc}
\inputencoding{ansinew}
\usepackage[OT1]{fontenc}
\usepackage{siunitx}
\usepackage{graphicx}
\usepackage{subeqn}
\usepackage{color}

\usepackage{hyperref}

\begin{document}

\title{Supramolecular structures in monohydroxy alcohols: Insights
  from shear-mechanical studies of a systematic series of octanol
  structural isomers} 

\author{Tina Hecksher} \author{Bo Jakobsen} \email{boj@dirac.ruc.dk}
\affiliation{DNRF centre ``Glass and Time'', IMFUFA, Department of
  Sciences, Roskilde University, Postbox 260, DK-4000 Roskilde,
  Denmark}

 
\begin{abstract}
  A recent study [Gainaru \textit{et al.}\ PRL.\, \textbf{112}, 098301
  (2014)] of two supercooled monohydroxy alcohols close to the
  glass-transition temperature showed that the Debye peak, thus far
  mainly observed in the electrical response, also has a mechanical
  signature. In this work, we apply broadband shear-mechanical
  spectroscopy to a systematic series of octanol structural isomers,
  $x$-methyl-3-heptanol (with $x$ ranging from 2 to 6). We find that
  the characteristics of the mechanical signature overall follow the
  systematic behavior observed in dielectric spectroscopy. However,
  the influence from the molecular structure is strikingly small in
  mechanics (compared to roughly a factor 100 increase in dielectric
  strength) and one isomer clearly does not conform to the general
  ordering. Finally, the mechanical data surprisingly indicate that
  the size of the supramolecular structures responsible for the Debye
  process is nearly unchanged in the series.
\end{abstract}



\maketitle 

Liquids forming supramolecular structures are currently receiving a
great deal of attention \cite{Lou2013, Griffin2013, Wang2014,
  Gainaru2014, Bohmer2014}.  In this communication, we present
shear-mechanical spectroscopy results on monohydroxy alcohols (i.e.,
alcohols containing only a single OH group).  A monohydroxy alcohol is
one of the simplest systems showing pronounced formation of
supramolecular structures and has been studied intensively in the past
mainly by dielectric spectroscopy (dating back 100 years), but also
by x-ray and neutron scattering, and calorimetry (see recent extensive
review by B\"ohmer \textit{et al.}\ [\onlinecite{Bohmer2014}] and
earlier compilation of classical results Ref.\ \onlinecite[Sec.\
IX-c.1]{Bottcher1980}).  Only few studies of the mechanical properties
of monohydroxy alcohols exist \cite{Lyon1956, Litovitz1963, Kono1966,
  Emery1978, Behrends2001, Jakobsen2008, Gainaru2014, Gainaru2014b},
but recent results \cite{Gainaru2014,Gainaru2014b} show that valuable
information about the supramolecular structures can be obtained by
mechanical spectroscopy.

Dielectric spectra for supercooled viscous monohydroxy alcohols close
to the glass-transition temperature show two pronounced anomalies:
Firstly, the dielectric constant (in the static limit) is different
from what is expected from the dipole moment of the molecules --- in
most cases it is much larger, but in some cases smaller. Secondly, the
dielectric spectrum shows a distinct feature at lower frequencies than
the main (alpha) structural relaxation. In many cases, this feature is
dominating, and --- contrary to the structural relaxation in most
molecular liquids --- is described by a single exponential
relaxation. This spectral feature thus follows the Debye-model
prediction, and is therefore often termed the ``Debye-peak'' or
``Debye-process''.

It is generally agreed (dating back to at least the seminal work by
Kirkwood \cite{Kirkwood1939} and later by W. Dannhauser
\cite{Dannhauser1968, Dannhauser1968b}) that these two dielectric
anomalies are related to supramolecular structures in the
liquid. The primary supramolecular structure is believed to be linear
(un-branched) chains, including rings and dimers. However, the
microscopic details of how these structures give rise to the observed
dynamic anomalies are far from settled, with different models
competing, e.g. Refs.\ \onlinecite{Gainaru2010b, Singh2012,
  Singh2013}.

Until recently, it was commonly accepted that the process leading
to the Debye-peak only showed a signature in dielectric spectroscopy,
what lead to many speculative arguments for why a signal was absent in
e.g.\ shear-mechanics and specific heat \cite{Huth2007, Lou2013,
  Bohmer2014}. This conventional belief was challenged, however, when
Gainaru \textit{et al.}\ [\onlinecite{Gainaru2014}] very recently
demonstrated the existence of a shear-mechanical low-frequency feature
in the spectra of two monohydroxy alcohols (2-ethyl-1-hexanol and
4-methyl-3-heptanol) close to the glass-transition
temperature. Applying two high resolution shear-mechanical techniques
enabled an identification of a low-frequency (i.e., sub alpha
peak-frequency) crossover from an intermediate behavior to a terminal
viscous behavior.  This spectral feature is interpreted to originate
from the same supramolecular structures responsible for the dielectric
Debye process, and is identical to what is observed for short chain
polymers \cite{Gray1977,Ferry1980}. Furthermore, it has been shown
that also NMR \cite{Gainaru2010b}, dynamic light scattering
\cite{Wang2014}, and quasi-elastic neutron scattering
\cite{Sillren2014} is sensitive to the Debye-process and similar
results have been seen in mechanical studies of \textit{supramolecular
  polymers} \cite{Lou2013} and high temperature alcohols
\cite{Behrends2001}. Altogether, this suggests that the Debye-process
is a ``normal'' relaxation phenomenon in the liquid, and not solely
connected to the dielectric properties, opening the route for many new
experiments which could provide valuable information on liquids
containing hydrogen bonded supramolecular structures.

\begin{figure}
  \includegraphics[scale=0.95]{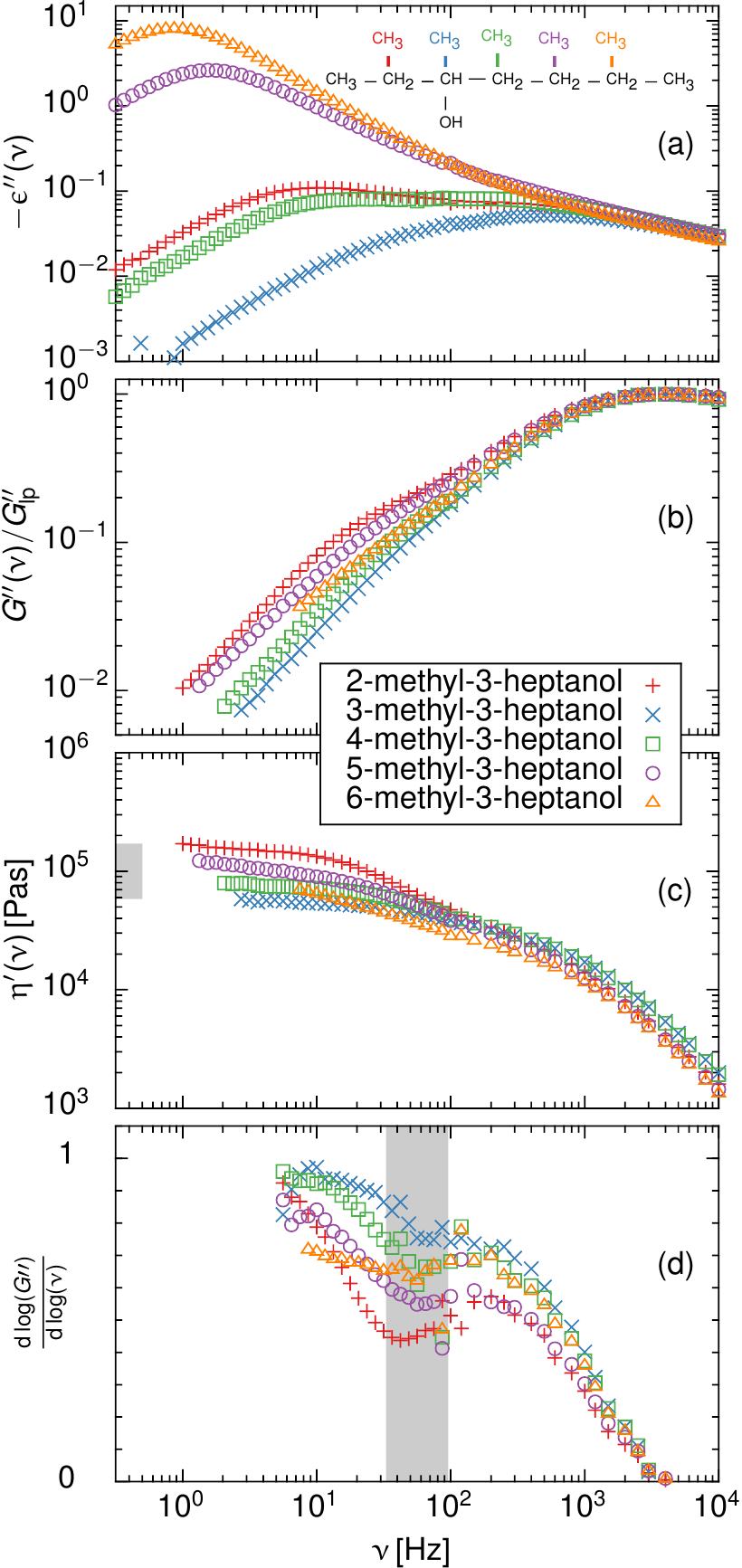}\\
  \caption{Comparison of the five studied octanol
    isomers, $x$M3H, at 182.5~\si{\kelvin} where the mechanical alpha
    loss peak, to a very good approximation, is at the same
    frequency. \textbf{a)} dielectric loss
    ($-\epsilon''(\nu)$). Included is an illustration of the molecular
    structure of the five isomers, color of the methyl group indicates
    the corresponding isomer and the backbone corresponds to the basic
    heptanol. \textbf{b)} Shear-mechanical loss ($G''(\nu)$)
    normalized by the value at the loss peak
    ($G''_\text{lp}$). \textbf{c)} Real part of shear viscosity
    ($\eta'(\nu)$). The gray region on the y-axis indicates the change
    in DC-viscosity observed between 3M3H and 2M3H.  \textbf{d)}
    Logarithmic derivative of the shear-mechanical loss, corresponding
    to the slope of the curves in subfigure b. The gray region
    indicates the expected change in crossover frequency between 3M3H
    and 2M3H from the observed change in viscosity (based on Eq.\
    \ref{eq:1}). The absolute position of the gray region on the
    x-axis is not predicted by Eq.\ \ref{eq:1} and is here chosen for
    best overlap with the observe change in crossover frequency.}
  \label{fig:1825k}
\end{figure}

In this work we present mechanical studies utilizing the broadband
technique applied in Ref.\ \onlinecite{Gainaru2014} on a series of
structural octanol isomers: $x$-methyl-3-heptanol ($x$M3H), with
$x$=2--6 (see Fig.\ \ref{fig:1825k}a). The aim is to investigate to
what extent the dielectric behavior carries over to the mechanical
response and to investigate what can be learned from mechanical
spectroscopy about the size and morphology of the supramolecular
structures. Our analysis is purely phenomenological and model-free,
taking the simplest possible approach hence avoiding unjustified
assumptions.

Studies of octanol isomers date back to the very early days of
dielectric measurements \cite{Smyth1929} with the work in the 1960's
of Dannhauser and Johari \cite{Dannhauser1968, Dannhauser1968b,
  Johari1969, Johari1969b} being the most thorough, but has also been
the subject of more recent investigations \cite{Singh2012, Singh2013,
  Bauer2013, Gainaru2014b}.

The $x$-methyl-3-heptanol system is interesting because the position
of the methyl group provides a tunable steric hindrance for the
hydrogen binding process, and thus a possible handle to changing the
supramolecular structures. In dielectric spectra (as shown in Fig.\
\ref{fig:1825k}a), this is manifested as a systematic increase in the
intensity of the Debye peak from a very weak, almost non-existing,
low-frequency feature in 3M3H to a dominant peak in 6M3H (this is
often represented as a variation in the Kirkwood factor
\cite{Kirkwood1939}, $g_\text{K}$). This systematic variation is
attributed to a change in size and morphology of the supramolecular
structures, from a ring dominated morphology (with small Debye-process
and $g_\text{K}<1$) to a chain dominated morphology (with dominant
Debye-process and $g_\text{K}>1$)
\cite{Dannhauser1968b,Singh2012,Singh2013}. However, it is hard, if
not impossible, to assign a quantitative microscopic model based only
on dielectric data, as stated by, e.g.,\ Dannhauser
\cite{Dannhauser1968b}: ``An attempt to make the forgoing conclusions
quantitative requires the introduction of a specific molecular
model''.

We measured the complex frequency-dependent shear modulus ($G(\nu) =
G'(\nu)+i G''(\nu)$) over a frequency range of
$10^{-3}$--$10^4~\si{\hertz}$ utilizing a unique broadband
shear-mechanical technique known as the ``Piezoelectric shear-modulus
gauge'' (PSG) technique \cite{Christensen1995}. The technique also
provides the complex, frequency-dependent shear viscosity as
$\eta(\nu) = G(\nu)/i2\pi\nu$ according to general response
theory. The conventional static viscosity, here denoted by $\eta_0$,
is defined as the zero-frequency limit of the real part of the
frequency-dependent viscosity, $\eta_0=\lim_{\nu\rightarrow 0}
\eta'(\nu)$.

The PSG technique is optimized for measuring on very stiff systems (in
the M$\si{\pascal}$-G$\si{\pascal}$ range) over a wide frequency
range. The shear-mechanical alpha relaxation of a glass-forming liquid
can thus be followed all the way to the glass-transition temperature.

\begin{figure}
  \includegraphics[scale=1]{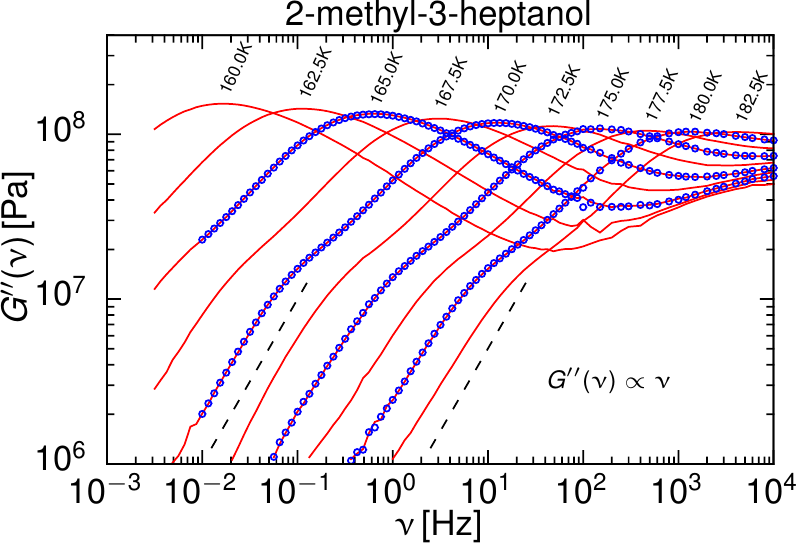}\\
  \caption{Shear-mechanical loss for
    2-methyl-3-heptanol as function of frequency ($G''(\nu)$) at the
    indicated temperatures. The main data, given in red, were taken
    stepping down in temperature, whereas the data shown as blue
    circles were taken reheating the sample, illustrating the
    reproducibility of the method. The dashed black lines indicate a
    slope of $1$ as expected for pure viscous behavior.  }
  \label{fig:raw}
\end{figure}

Corresponding dielectric spectra were measured in tandem with the
mechanical data under identical experimental conditions to allow for a
direct comparison.

The five octanol isomers were used as received \footnote{Details of
  samples used: \textbf{2M3H} from Alfa Aesar (90\%
  purity). \textbf{3M3H,4M3H,5M3H} from Sigma Aldric.  \textbf{6M3H}
  from TCI. 4M3H shear mechanical data reproduce the results from the
  study published in \onlinecite{Gainaru2014}, the dielectric data was
  taken during that study.}. For each sample a broad temperature and
frequency scan was made down to temperatures where the mechanical
alpha loss peak is at approximately
$10^{-1}~\si{\hertz}$.\footnote{The presented data can be obtained
  from the “Glass and Time: Data repository,” see
  \url{http://glass.ruc.dk/data}.}

Figure \ref{fig:raw} shows the imaginary part of the shear modulus for
2M3H at different temperatures.  On the low-frequency side of the
alpha peak, a clear crossover from one intermediate power law to a
terminal power law of $\nu^1$ --- characteristic of viscous behavior
--- is seen for all temperatures. We refer to this as \textit{the
  crossover to pure viscous behavior}. This behavior complies with the
observations in Refs.\ \onlinecite{Gainaru2014,Gainaru2014b},
interpreted as short chain polymer-like dynamics.

\begin{figure}
  \includegraphics[]{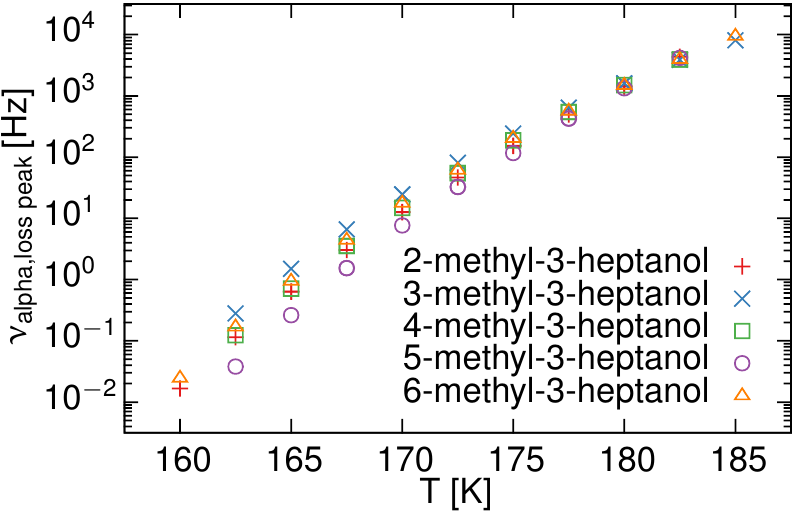}\\
  \caption{Shear-mechanical loss-peak frequency as
    function of temperature for the five studied octanol isomers.}
  \label{fig:lpf}
\end{figure}

The loss-peak frequencies determined as function of temperature for
all the isomers are shown on Fig.\ \ref{fig:lpf}. The direct access to
the characteristic time scale of the alpha relaxation is one of the
big advantages of using broadband mechanical spectroscopy on
monohydroxy alcohols. The alpha time scale in the dielectric data can
only be estimated through somewhat questionable model fits due to the
presence of the Debye peak.

The alpha time scale of the different isomers follow each other rather
closely. At 182.5~\si{\kelvin} the alpha time scales are practically
identical, with the alpha loss peak at 4~k\si{\hertz}. At the lowest
temperature where data exists for all 5 liquids (162.5~\si{\kelvin})
they differ less than a factor of 10 after having changed $5$ orders
of magnitude.  No clear trend, with respect to molecular
configuration, in the ordering of alpha time scales at low
temperatures is observed.

In the following we will focus on a close analysis of the data set
from 182.5~\si{\kelvin}, where a unique possibility exists for
studying the influence of the molecular geometry on the mechanical
crossover to pure viscous behavior, while keeping the alpha relaxation
time and temperature constant.

Figure \ref{fig:1825k}a and \ref{fig:1825k}b show the dielectric and
shear-mechanical loss for all five liquids at 182.5~\si{\kelvin}. The
dielectric data, qualitatively, reproduce results from the literature
\cite{Dannhauser1968,Dannhauser1968b,Singh2013,Singh2012} showing that
the Debye peak intensifies and shifts to lower frequencies when the
methyl group is moved away from the hydroxyl group while the alpha
peak position is unchanged, i.e., a strong dependence of the Debye
process on the molecular geometry.

Comparing Fig.\ \ref{fig:1825k}a and \ref{fig:1825k}b, the alpha
relaxation in the mechanical data is shifted to higher frequencies
than in the dielectric data. It is a common feature for molecular
viscous liquids that the shear-mechanical alpha relaxation is shifted
to higher frequencies compared to dielectric spectroscopy (see, e.g.,
Refs.  \onlinecite{Jakobsen2005, Niss2005, Jakobsen2008, BetaShear}
and references therein). In addition, we now observe that the
crossover to pure viscous behavior is also shifted to higher
frequencies compared to the dielectric Debye peak. Thus, the overall
shear mechanical relaxation is faster than the dielectric, not just
the alpha relaxation. That different response functions relax on
different time scales is no surprise \cite{Jakobsen2012}, but it is
interesting that all visible processes are faster in shear mechanics
than in dielectric relaxation.

The shape of the shear-mechanical relaxation on the low-frequency side
of the alpha peak depends on the molecular geometry. The separation
between the alpha relaxation and the crossover frequency mimics the
order from the dielectric spectra. However, it is clear that 2M3H,
which from a dielectric point of view is close to 4M3H (as one would
expect based on the molecular geometry), behaves much more like 5M3H
in the mechanics with a surprisingly pronounced and temporally
separated mechanical crossover signal.

Figure \ref{fig:1825k}c shows the calculated frequency dependent
viscosity for the 5 liquids. Again, we observe increasing viscosity
with increasing separation of the methyl and hydroxyl group. But also
in this signal 2M3H is contradicting the general trend by having a
relatively large viscosity compared to 4M3H and even 5M3H.

Figure \ref{fig:1825k}d show the logarithmic derivative of the
mechanical loss, $\frac{d \log (G''(\nu))}{d \log(\nu)}$, which is
simply the slope of the curves in Fig.\ \ref{fig:1825k}b. This
representation of the data is specifically well suited to reveal power
law behavior in the spectra and thus shows the crossover to the
terminal viscous behavior much clearer than the raw data. As expected
from the raw data, the curves display a two step behavior going down
in frequency: from 0 (at the alpha peak frequency) approaching 1 at low
frequencies (equivalent to the final viscous behavior where
$G''(\nu)\propto\nu$). In the one extreme case, 3M3H, almost no
intermediate plateau is observed due to the very small separation
between the alpha relaxation and the terminal viscous behavior. In the
other extreme case, 6M3H, the terminal behavior is outside the
accessible range of the technique and only the intermediate plateau is
observed. This is also seen in the viscosity data (Fig.\
\ref{fig:1825k}c) where the plateau --- signaling the onset of pure
viscous flow --- is not reached within the frequency window and
resolution available here.

Comparing the mechanical and dielectric data, it is striking that the
difference between the isomers is relatively small in the mechanical
data compared to the huge difference in dielectric strength. Likewise
the influence on the temporal separation between the alpha and Debye
relaxation seems smaller in the mechanical signal. The striking
increase in dielectric constant with change in molecular geometry can
be attributed to a change in both size and morphology of the
supramolecular structures, but dielectric spectroscopy is not able to
directly distinguish between the two effects. In the following, we
investigate the apparent size of the supramolecular structures
determined from the mechanical data.

The relation between effective volume, $V_\text{eff}$, and
characteristic time scale $\tau$ of an object in a viscous environment
(with viscosity $\eta_0$) at a given temperature is in general given
by:
\begin{eqnarray}
  \label{eq:tau_gen}
  \tau \propto \frac{\eta_0 V_{\text{eff}}}{k_\text{B} T}.
\end{eqnarray}
This result can be obtained from simple dimension analysis, and
several models predict such a relation, e.g., the Debye model
\cite{Debye1929} and the Rouse model prediction for the slowest Rouse
mode in polymer melts \cite{Ferry1980}
($\tau_\text{Rouse}=\frac{6\eta_0 M}{\pi^2 \rho R T}$, where $M$ is
the molecular weight and $\rho$ is the density of the polymer). In
Ref.\ \onlinecite{Gainaru2014} the Rouse prediction was used to
estimate the length of the expected chain-like structures in
2-ethyl-1-hexanol and a reasonable result was found with a chain
consisting of $\approx 9$ alcohol molecules.

In the following we apply the general formulation in Eq.\
\ref{eq:tau_gen} to evaluate whether the mechanical data suggest a
growing size of the supramolecular structures in the $x$M3H octanol
isomer series. 

Equation \ref{eq:tau_gen} predicts (assuming a constant geometrical
prefactor and constant temperature) that for any pair of isomers, $X$
and $Y$, we have:
\begin{eqnarray}
  \label{eq:1}
  \log\frac{\tau_{\text{cross,X}}}{\tau_{\text{cross,Y}}} =
  \log\frac{\eta_{0,\text{X}}}{\eta_{0,\text{Y}}} + 
  \log\frac{V_{\text{eff,X}}}{V_{\text{eff,Y}}},
\end{eqnarray}
where $\tau_{\text{cross}}$ is the time for the crossover to viscous
behavior.  A growing size should thus result in a $\tau_{\text{cross}}$
which changes more than expected from just the changes in viscosity.

For the present case the small separation between the pure viscous
behavior and the alpha relaxation makes it difficult to estimate
$\tau_{\text{cross}}$ without unjustified model assumptions
\cite{Roland2004, Gainaru2014b}. In order to overcome this problem, we
calculated the expected change in $\tau_{\text{cross}}$ based purely
on viscosity increase (i.e., assuming $V_{\text{eff,X}} /
V_{\text{eff,Y}}=1$). 

The viscosity change between the two extremes where viscosity is well
defined (3M3H and 2M3H) is a factor three (indicated in Fig.\
\ref{fig:1825k}c by the grey shaded area on the $y$-axis).  The
corresponding expected change in $\tau_{\text{cross}}$ has been
indicated by the grey shaded area in Fig.\ \ref{fig:1825k}d.
Surprisingly the observed change in crossover frequency between the
four liquids 2M3H -- 5M3H is clearly captured in the shaded area
(independent on how the crossover frequency is defined).

This shows that the volume of the supramolecular structure does not
change significantly between the four isomers (under the assumption of
a constant geometry factor in Eq.\ \ref{eq:tau_gen}), as the observed
change in separation between the alpha relaxation and the crossover to
terminal viscous behavior can be explained purely by the increase in
viscosity.

Furthermore, there are good reasons to assume that the isomers have
the same density at the measuring temperature; they have the same
density at room temperature and are expected to have roughly the same
expansion coefficient \cite{Dannhauser1968b}. Under this assumption,
the result above directly translates to an unchanged number of
molecules participating in one supramolecular structure.  This finding
supports the idea that the dominating mechanism for the extremely
large increase in the dielectric Debye strength is a change in
geometry of the supramolecular structures, as was suggested from
Kirkwood analysis of dielectric data
\cite{Dannhauser1968b,Singh2012,Singh2013}.

\begin{figure}
  \includegraphics[]{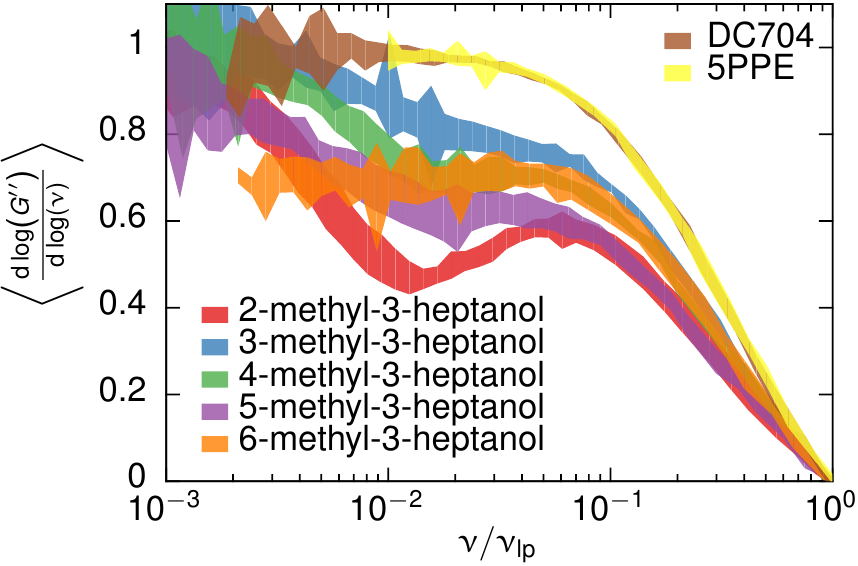}
  \caption{Mean logarithmic derivative of the shear-mechanical loss
    for the 5 studied octanol isomers and the two van der Walls
    bounded molecular liquids tetraphenyl-tetramethyl-trisiloxane
    (DC704) and 5-phenyl-4-ether (data from
    \onlinecite{Hecksher2013}). For each substance the mean is taken
    over all available data sets after rebining the data on a common
    scaled frequency axis. The width of the curve indicates mean $\pm$
    one standard derivation. The standard derivation is zero if time
    temperature superposition holds, leading to line like curves.}
  \label{fig:slope}
\end{figure}

In order to show the generality of the observations from the data set
taken at $182.5\si{\kelvin}$ the logarithmic derivative data at all
temperatures were averaged using a scaled frequency axis
($\log_{10}(\nu/\nu_{lp})$).  Figure \ref{fig:slope} shows the result
of this procedure and also includes data for two van der Waals bonded
liquids for comparison. The generality of the results is clear, which
also means that the overall shape of the mechanical relaxation curves is
not too temperature dependent (otherwise the curves would be very
smeared out).

In general, the systematic behavior found in dielectric spectroscopy
of these liquids is not as clear in the mechanical response. The
separation between the mechanical alpha and Debye process (and
corresponding viscosity contribution) increases with increasing
separation of the methyl group and the hydroxyl group. However, 2M3H
is an exception; in dielectric spectroscopy it is close to 4M3H with a
small Debye peak as expected, but in mechanics it has a very large and
well separated Debye signal somewhere between 5M3H and 6M3H. The
temperature dependence of the mechanical alpha time scale (or,
equivalently, the fragility index) does not follow a specific ordering
in contrast to what was found for the dielectric alpha
\cite{Singh2013}.

The observation of a clear mechanical signal from the supramolecular
structures establishes that the process leading to the dielectric
Debye peak is not exclusively related to dielectric properties, but a
general relaxation phenomenon in monohydroxy alcohols. Therefore one
would expect that an equivalent signal exists in other response
functions, e.g., specific heat. The lack of present observation could
very well be due to limited resolution as was the case of earlier
mechanical investigations (e.g., in Ref.\ \onlinecite{Jakobsen2008}).

The mechanical data suggests that the size of the supramolecular
structures when going from 3M3H -- 4M3H -- 5M3H -- 2M3H is unchanged,
in contrast to a factor 100 increase in dielectric strength. We
conclude that the change in dielectric strength is dominated by
changes in geometry of the supramolecular structures and not a growing
chain length.

\begin{acknowledgments}
  The centre for viscous liquid dynamics ``Glass and Time'' is
  sponsored by the Danish National Research Foundation via grant
  DNRF61
\end{acknowledgments}


%

\end{document}